*CORRESPONDENCE
Tamal Basak,
tamalbasak@gmail.com

SPECIALTY SECTION
This article was submitted to Atmosphere and Climate, a section of the journal Frontiers in Environmental Science

# On the altitude profile of lower ionospheric D-region response time delay during solar flares

Sayak Chakraborty[1], Rakhi Paul[2] and Tamal Basak[3]*

[1]Indian Centre for Space Physics, Kolkata, India, [2]Department of Physics, Bethune College, Kolkata, India, [3]Department of Physics, Amity University Kolkata, Kolkata, India

The D-region ionosphere is sluggish in nature while responding to the incoming solar radiation. We study the altitude ($h$) profile of mid-latitude D-region response time delay ($\Delta t$) during three chosen solar flares, namely, C5.2, M5.2, and X2.2 classes. We solve "electron continuity equation" numerically at each and every $h$ to obtain $\Delta t$-$h$ profile. We investigate the 1) latitudinal variation (over both Northern and Southern hemispheres) and 2) seasonal variation (throughout the year) of $\Delta t$-$h$ profiles of each of these solar flares separately. We go over noteworthy variations of $\Delta t$-$h$ profile for both 1) and 2) with reasonably different results over different hemispheres. Also, we study some contrasting behaviours of $\Delta t$-$h$ profile of X2.2 class for both 1) and 2) in comparison to C5.2 or M5.2 classes. We conclude the outcome with possible explanations.

KEYWORDS

**d-region ionosphere, x-ray solar flare, response time delay, latitudinal variation, seasonal variation**

## 1 Introduction

The ionospheric free electron content gets enhanced due to absorption of the incoming solar irradiation Midya and Chattopadhyay (1996); McRae and Thomson (2004); Kolarski et al. (2011); Berdermann et al. (2018); Chakrabarti et al. (2018); Gavrilov et al. (2019); Chakraborty and Basak (2022) etc.,. But, the response of the ionosphere is not instantaneous to such incoming solar irradiation, rather, it measurably lags Jones (1971); Valnicek and Ranzinger (1972); Basak and Chakrabarti (2013); Palit et al. (2015); Chakraborty and Basak (2020) etc., Appleton (1953) described the term "sluggishness" as the time delay between peak irradiance at solar noon and the resulting peak in the ionospheric electron density ($N_e$). Solar flare is one of such prominent solar events which causes sudden ionospheric enhancement (SIE) and ionospheric sluggishness is also observed. Degree of such ionospheric perturbation varies from layer to layer significantly. Also, depending upon the strength of the solar flare, the degree of ionisation and sluggish nature of the ionosphere changes. Thomson and Clilverd (2001), McRae and Thomson (2004), Žigman et al. (2007), Basak and Chakrabarti (2013), Palit et al. (2015), Chakrabarti et al. (2018), Hayes et al. (2021), Nina (2022) and many others reported a finite and measurable enhancement of D-region ionospheric electron density ($N_e(t)$) during solar flares by studying the sub-ionospheric radio signal propagation effect. By absorbing the incoming solar irradiation during a solar flare, $N_e(t)$ starts increasing and reaches at a maximum value ($N_{e,max}$). Thereafter, it decays gradually. The net electron number density ($N_e(t)$) of the ionosphere is a result of the balance between the ionisation and electron loss effects. In the mid-latitude D-region, the electron recombination process dominates the electron loss effect Friedrich et al. (2004); Nina et al. (2012). Following Chakraborty and Basak (2020), we define the effective response time delay of D-region ionosphere ($\Delta t$) as the time difference between the maxima of solar x-ray flux ($\phi_{max}$) and maxima of corresponding electron density ($N_{e,max}$) during a given solar flare. It is expressed as,

$$\Delta t = t_{\phi_{max}} - t_{N_{e,max}} \tag{1}$$

Where, $t_{\phi_{max}}$ and $t_{N_{e,max}}$ are the timings of occurrences of $\phi_{max}$ and $N_{e,max}$ respectively. Le et al. (2007), Chakraborty and Basak (2020) reported the dependency of $\Delta t$ on solar zenith angle ($\chi$). Solving the D-region 'electron continuity equation' in presence of three solar flares of different classes, Chakraborty and Basak (2020) reported significant seasonal and latitudinal variation of $\Delta t$ at a fixed altitude of 74 km.

The degree of ionisation varies with the ionospheric altitude due to the variation in the amount of solar energy deposition and as a result, the $\Delta t$ must have a significant altitude dependency as well. Appleton (1953) reported that $\Delta t$ for various ionospheric layers should be of the order of $\frac{1}{2\alpha N_e}$, where, $\alpha$ is the standard "recombination coefficient". Mitra (1952) opted a theoretical approach starting from the considerations of 'Chapman region' and reported an altitude profile of negative ions to electron number density ratio ($\lambda$, a dimensionless number) and $N_e(t)$ under ambient ionospheric conditions for D-region ionosphere. In this regard, they assumed an idealised temperature distribution (i.e., the scale height ($H$) profile) adopted by the National Advisory Committee for Aeronautics (NACA), United States . Moreover, they reported a significant relation of solar zenith angle ($\chi$) profile with $N_e(t)$. Parthasarathy and Rai (1966) deduced the altitude profile of 'effective recombination coefficient ($\alpha_{eff}$)' along with $N_e(t)$ and electron production rate ($q(t)$) of D-region ionosphere using multiple frequency radio wave absorption data. Palit et al. (2015) studied the sub-ionospheric VLF radio signal propagation effect and computed the ion densities. They developed an altitude ($h$) profile of $\Delta t$ for three solar flares of different classes using numerical techniques. They reported the same for $N_{e,max}$ also. They did it for D-region over a specific geographical area and on the day of solar flare occurrence. Nina and Čadež (2021) worked on the electron loss rate resulting from recombination processes in D-region ionosphere. They looked at the altitude dependency of the respective plasma electron loss rate and its spatial (and temporal) derivatives during a solar flare. Different types chemical reactions with their different reaction time rates ($\tau$) occur at different altitudes during a solar flare. Mitra and Rowe (1972) reported some of such chemical reactions along with their $\tau$'s at 70 km and 80 km altitudes respectively. They reported the notably different $\tau$'s in case of cluster ions in these two different altitudes. These kind of works lead us to investigate the overall altitude ($h$) dependency of $\Delta t$ during solar flares in a more rigorous fashion.

Hayes et al. (2021) analysed $\Delta t$ measured from VLF amplitude observations in association with soft (and hard) X-ray profiles as taken from GOES-15 satellite for 334 solar flares. They analysed the VLF signal for the propagation path from NAA (Maine, United States) to Offaly, Ireland. They considered the solar flares occurred in between 2012 and 2018. 261C-class, 71M-class and 2 X-classes of solar flares are included in their study. Their analysis was limited to the study of $\Delta t$ at the sub-ionospheric VLF signal reflection height in D-region. We compare $\Delta t$ values as observed by VLF signal in Hayes et al. (2021) with the same obtained from our methodology over similar geographical coordinates. We found that the values computed by us are in the similar order with the same reported by Hayes et al. (2021). For a typical C5 class solar flare, the $\Delta t$ reported in Hayes et al. (2021) is within the overall range ~ 60–300 s. We get the $\Delta t$ values within the similar range for the solar flares of similar classes. The average $\Delta t$ value for typical C5 class solar flares, as measured by Chakraborty et al. (2022) is ~ 150–250 s for 60°N latitude. These results also are in the range of the outcomes of our analysis for the similar D-region altitude. A work similar to Hayes et al. (2021) was performed by Chakraborty et al. (2022). They computed $\Delta t$ values numerically for 455 solar flares over the mid-latitude D-region occurred during January-February of the years from 2011 to 2016. The $\Delta t$ values computed by us for C, M and X-classes of flares are in the similar range of $\Delta t$ values computed by Hayes et al. (2021), Chakraborty et al. (2022) etc., Hayes et al., 2021 noted a few negative $\Delta t$ values, whereas, we have no such observation. Because, negative $\Delta t$ values are possible due to the presence of the day-night terminator near the VLF signal propagation path. Our study is based on "D-region electron continuity equation" and it does not include VLF based $\Delta t$ measurements.

In this article, we take a generalised approach regarding the analysis of $\Delta t$. We consider the latitudinal and seasonal variations of the altitude ($h$) profile of $\Delta t$ in presence of the solar x-ray irradiance during the solar flares. We assume the altitude profiles of relevant D-region parameters and numerically solve the "electron continuity equation" (Whitten and Poppoff (1961); Rowe et al. (1970); Ananthakrishnan et al. (1973); Žigman et al. (2007); Basak and Chakrabarti (2013); Palit et al. (2016); Palit et al. (2018); Nina et al. (2018); Chakraborty and Basak (2020) etc.) (Eq. 2) for a set of three solar flares each of different classes (see Table 1). We compute $N_e(t)$ profile and subsequently the $\Delta t$'s from Eq. 1. We perform a thorough comparison of the $\Delta t$-$h$ profiles computed over each of the chosen latitudes on each day of the year ($DoY$). We present a robust and systematic nature of $\Delta t$-$h$ profile with possible physical explanations.

TABLE 1 Details of solar flares.

| Class of solar flare | Date of occurrence | $t_{\phi_{max}}$ (hh:mm:ss in UT) | Longitude, where, $t_{\phi_{max}}$ = 12 h (local time) |
|---|---|---|---|
| C5.2 | 24.02.2014 | 21: 37: 12 | $215^{\circ}37'40.8''$ |
| M5.2 | 04.02.2014 | 04: 00: 36 | $119^{\circ}50'24''$ |
| X2.2 | 15.02.2011 | 01: 57: 00 | $150^{\circ}47'0.96''$ |

## 2 Methodology

The lower ionospheric D-region electron density variation at any given altitude ($h$) is primarily governed by "electron continuity equation". The generalised form of it is as follows,

$$\frac{dN_e(t)}{dt} = \frac{q(t)}{1+\lambda(t)} - \frac{N_e(t)}{1+\lambda(t)}\frac{d\lambda(t)}{dt} - \alpha_{eff} N_e^2(t). \quad (2)$$

We adopt the numerical iterative method as described in Chakraborty and Basak (2020) to solve Eq. 2. For further analysis, we consider the C5.2, M5.2 and X2.2 classes of solar flares (see Table 1). We use the temporal profile of soft x-ray ($\phi(t)$) from GOES-15 satellite recorded during these solar flares (having the temporal resolution marginally greater than 2 s). Therefore, we solve Eq. 2 to generate the respective $N_e(t)$ profiles. Further, we calculate $\Delta t$ for these solar flares using Eq. 1. Chakraborty and Basak (2020) followed a similar methodology to compute the latitudinal and seasonal dependency of $N_e(t)$ and $\Delta t$. But, those analysis were done at a specific D-region altitude of 74 km. Here, we try to look at it with a much broader spectrum by investigating the entire altitude profile of $\Delta t$. In spite of being the thinnest layer in the ionosphere, the chemical dynamics of D-region is the most complicated one and has a strong altitude variation within the available altitude span of this region. As a result, $q(t)$, $\lambda(t)$ and $\alpha_{eff}$ in Eq. 2 reportedly have notable altitude dependency. So, in order to compute the $\Delta t$-$h$ profile, we use the $h$-dependencies of the mentioned parameters. We adopt the standard profiles of $\lambda(t)$ and $\alpha_{eff}$ for different classes of solar flares from Palit et al. (2015), where, $\alpha_{eff}$ has a tendency of decaying with $h$ ranging from 60 km to 80 km for all three classes of flares across mid-latitude region. At 60 km for a C-class flare chosen by them, the value of $\alpha_{eff}$ is $3.25 \times 10^{-11}$ m$^3$ s$^{-1}$ and at 80 km it is $1.4 \times 10^{-11}$ m$^3$ s$^{-1}$. In lower altitude range (60–65 km), all the three classes of flares share almost same $\alpha_{eff}$ values (within a range of nearly $3.09 \times 10^{-11}$–$3.25 \times 10^{-11}$ m$^3$ s$^{-1}$) and after 65 km, $\alpha_{eff}$ ($\sim 1.6 \times 10^{-11}$ m$^3$ sec$^{-1}$ at 80 km) increases at a faster rate than that of X-class flare ($\sim 1.4 \times 10^{-12}$ m$^3$ sec$^{-1}$ at 80 km). According to Palit et al. (2015), the $\lambda(t)$ has a decreasing nature with altitude ($h$) which is similar to $\alpha_{eff}$ ranging from 60 km to 80 km for all three classes of flares. The value of $\lambda(t)$ at 60 km for the C-class flare is nearly 27 and at 80 km, it is 0.5. $\lambda(t)$ decays sharply in lower altitude region (60–65 km) for C-class flare. At higher altitudes (75–80 km), $\lambda(t)$ becomes nearly "zero". $\lambda(t)$ values for the C-class flares are higher than the same for X-class flares in the lower altitude. For example, $\lambda(t)$'s at 60 km are 27 and 13 for C-class and X-class respectively. According to Ratcliffe (1973), Budden (1988), Žigman et al. (2007), Basak and Chakrabarti (2013), Chakraborty and Basak (2020), the expression of $q(t)$ is as follows,

$$q(t) = \frac{C\phi(t)}{eH} cos\,\chi(t), \quad (3)$$

where, '$C$' (=$1.838 \times 10^{17}$ J$^{-1}$) is the number of ions produced by the absorption of unit amount of radiation. As reported by Mitra (1952), Mitra (1992), Basak and Chakrabarti (2013), $H$ is proportional to ionospheric equilibrium temperature ($T$). The standard altitude profile of $T$ was reported by Mitra (1952) and we adopted the same to establish the altitude profile of $H$ (and subsequently $q(t)$) using Eq. 3. The standard $H$ profile below 100 km altitude under isotherm condition was adopted by Mitra (1952) from National Advisory Committee for Aeronautics (NACA). Mitra (1952) claimed that although $T$ may have dependency on $\chi(t)$, but there should not be much deviation from NACA's standard $T$ distribution. The $T$ decreases with increasing D-region altitude and its value ranges in between 350 and 240 K. Palit et al. (2013) reported a strong altitude and $\chi(t)$ dependency of $q(t)$. Some of the values of $q_{\max}$ as obtained from Eq. 3 are $2 \times 10^7$ m$^{-3}$ s$^{-1}$ (for C5.2, N60, $h$ = 60 km), $3.02 \times 10^8$ m$^{-3}$ s$^{-1}$ (for M5.2, N45, $h$ = 70 km), $2.43 \times 10^9$ m$^{-3}$ s$^{-1}$ (for X2.2, S30, $h$ = 80 km) etc., Such variation in $q_{\max}$ with $h$ and latitude for different classes of flares leads to generalised profiles of $N_e(t)$ and $\Delta t$. We compute $N_e(t)$-$h$ and $\Delta t$-$h$ profiles ranging from $h$ = 60–80 km using Eq. 2 (Figure 1). We obtain the latitudinal and seasonal variations of $\Delta t$-$h$ profiles by supplying the suitable values of $\chi(t)$ to Eq. 3 from https://www.esrl.noaa.gov. We purposefully restrict our computation across D-region over mid-latitudinal area for both the hemispheres as the ionospheric effects of precipitation of charge particles (Vontrat-Reberac et al. (2001)) across high latitude are not included in Eq. 2. Hence, we study the latitude dependency $\Delta t$-$h$ profile within the latitude range from 30° to 60° in both the hemispheres (Figure 2). Thereafter, for studying the seasonal variation, we assume that the C5.2, M5.2, and X2.2 classes repeat themselves on each day of the year ($DoY$). For this analysis, we choose the D-region above six different latitudes, namely, 30°N, 45°N, 60°N, 30°S, 45°S, and

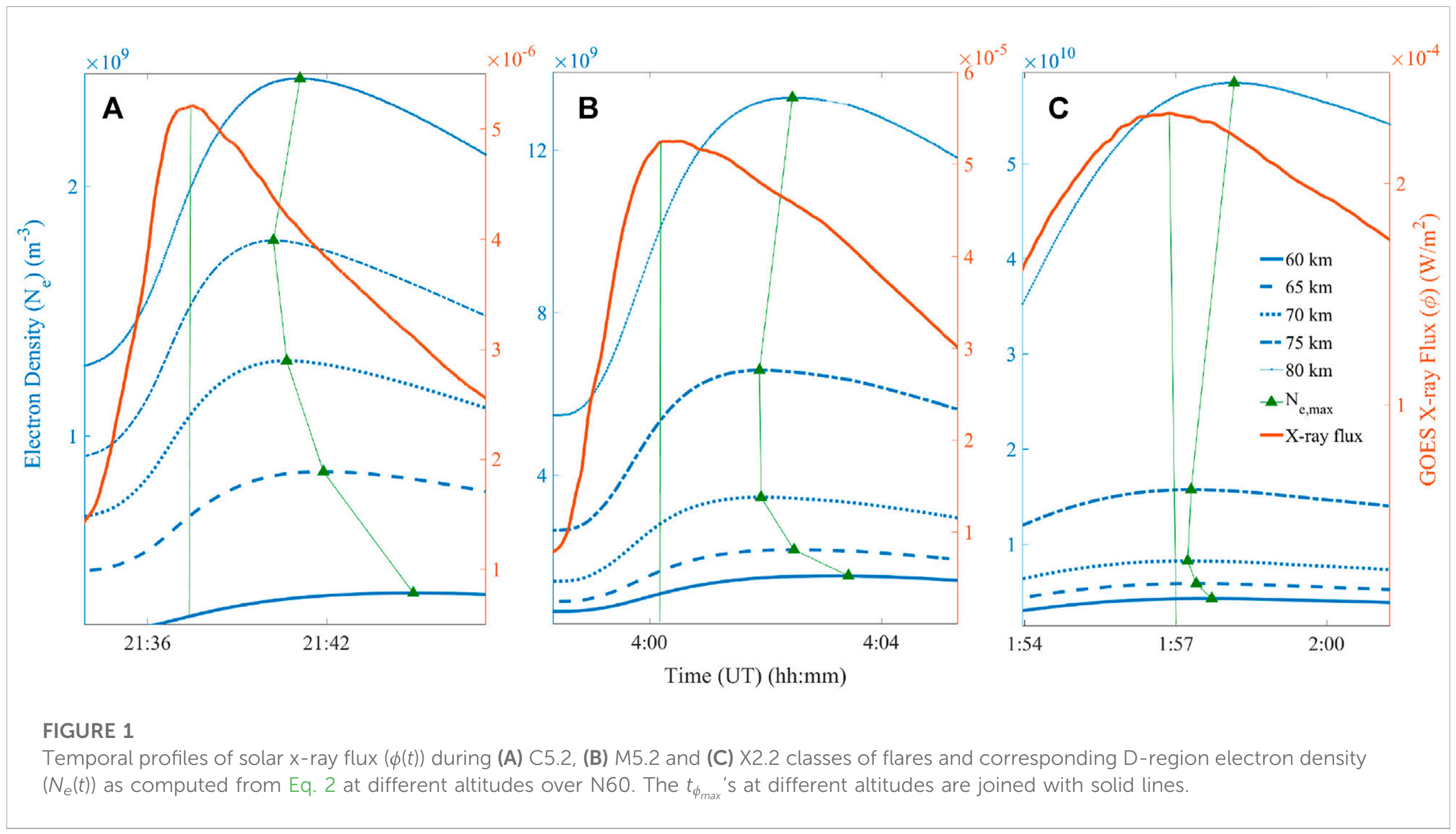

FIGURE 1
Temporal profiles of solar x-ray flux ($\phi(t)$) during **(A)** C5.2, **(B)** M5.2 and **(C)** X2.2 classes of flares and corresponding D-region electron density ($N_e(t)$) as computed from Eq. 2 at different altitudes over N60. The $t_{\phi_{max}}$'s at different altitudes are joined with solid lines.

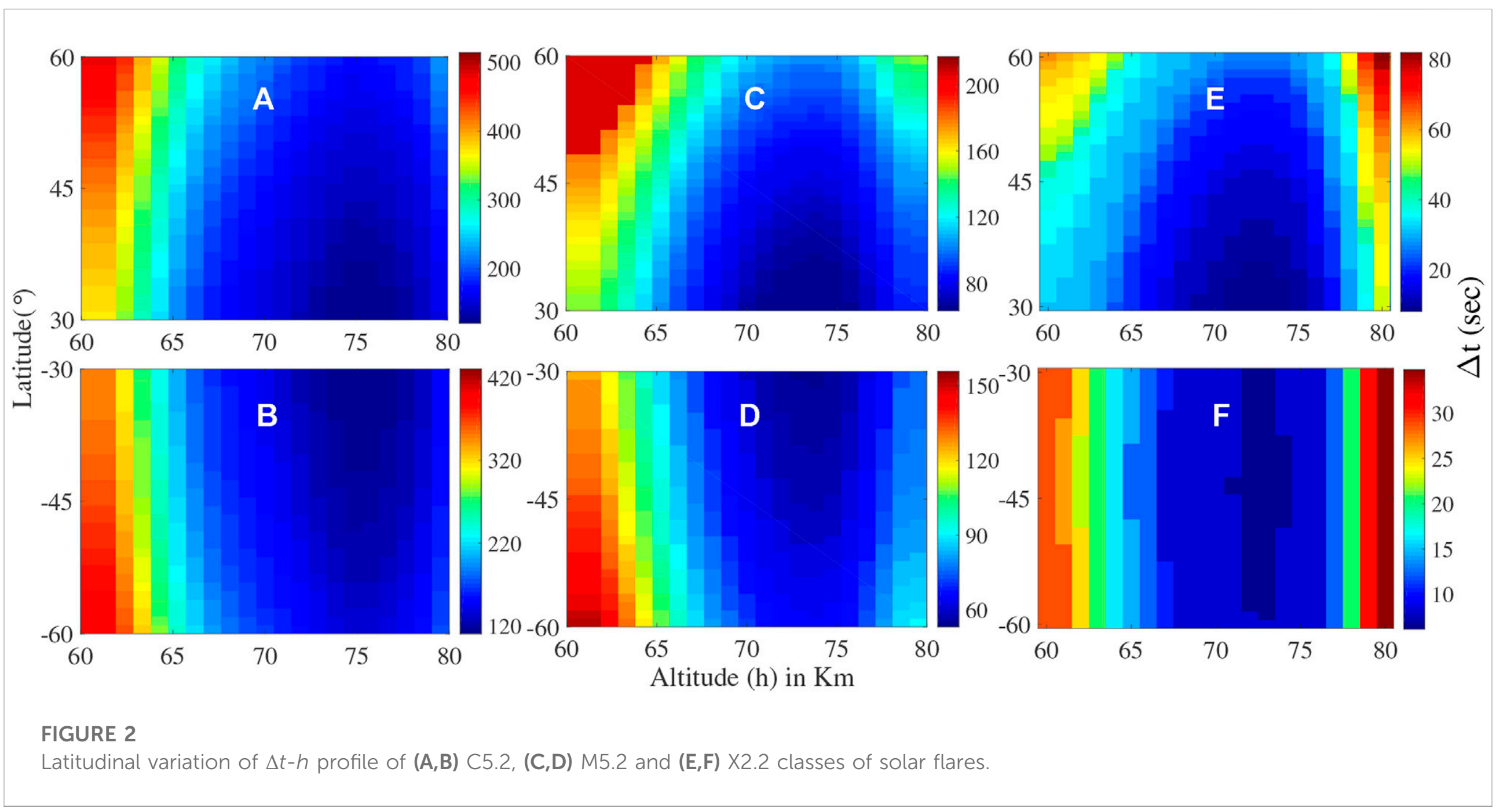

FIGURE 2
Latitudinal variation of $\Delta t$-$h$ profile of **(A,B)** C5.2, **(C,D)** M5.2 and **(E,F)** X2.2 classes of solar flares.

60°S (Figures 1, 3, 4). We would call them as N30, N45, N60, S30, S45, and S60 respectively from now on. 30°N, 60°N, 30°S and 60°S are the boundaries and 45°S and 45°N are the middles of mid-latitude regions. Here, we check the $\Delta t$-$h$ profile starting from the first day ($DoY = 1$) to the last day ($DoY = 365$) of the year (Figures 3, 4). The data resolutions for latitude, $h$ and $DoY$ are taken as 1°, 1 km and 1 day respectively. For the entire analysis, we choose the D-region only above those specific longitudes respectively for all three solar flares, where, the $t_{\phi_{max}} = 12$ h (local longitude time) (see Table 1).

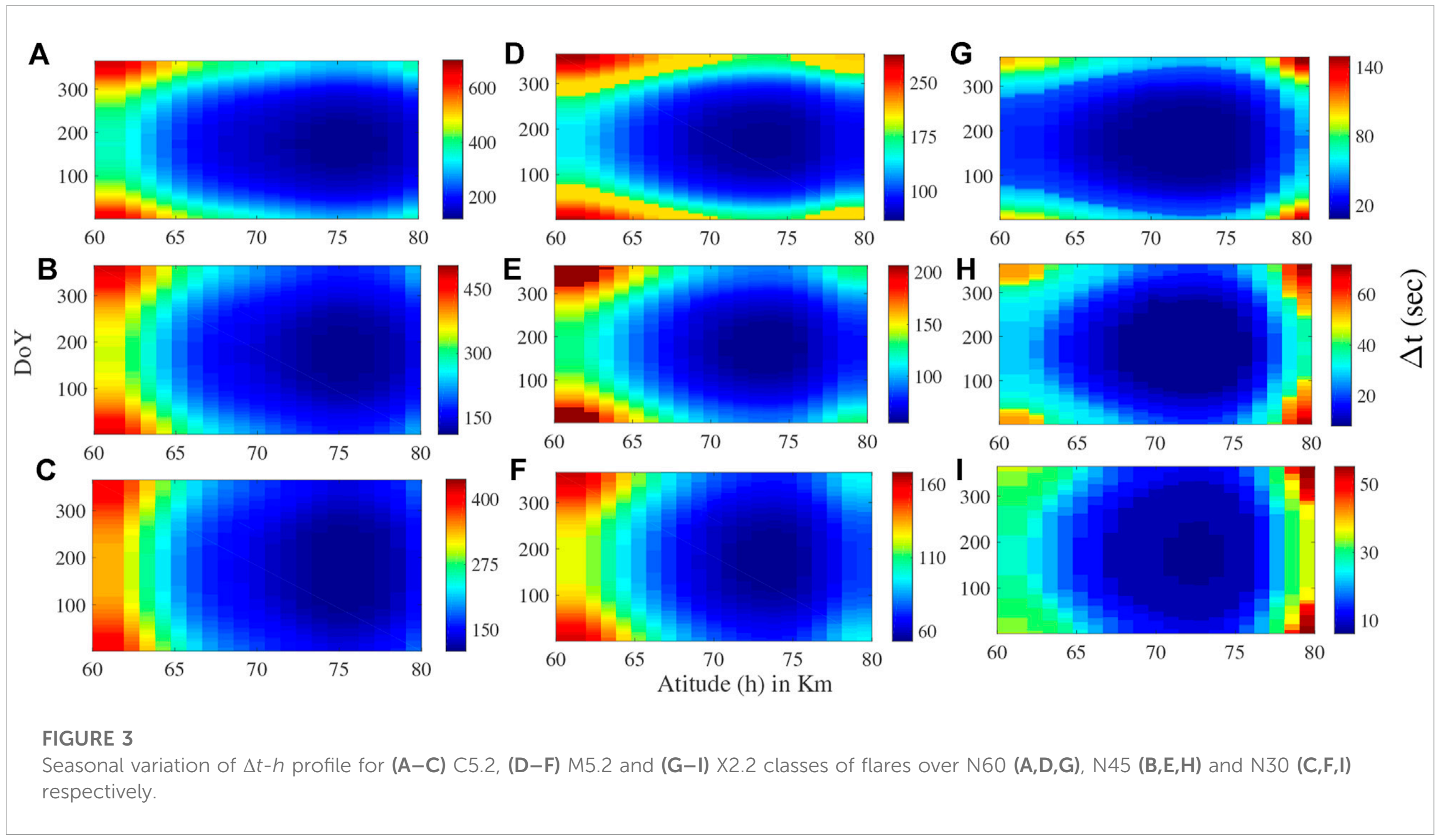

FIGURE 3
Seasonal variation of $\Delta t$-$h$ profile for **(A–C)** C5.2, **(D–F)** M5.2 and **(G–I)** X2.2 classes of flares over N60 **(A,D,G)**, N45 **(B,E,H)** and N30 **(C,F,I)** respectively.

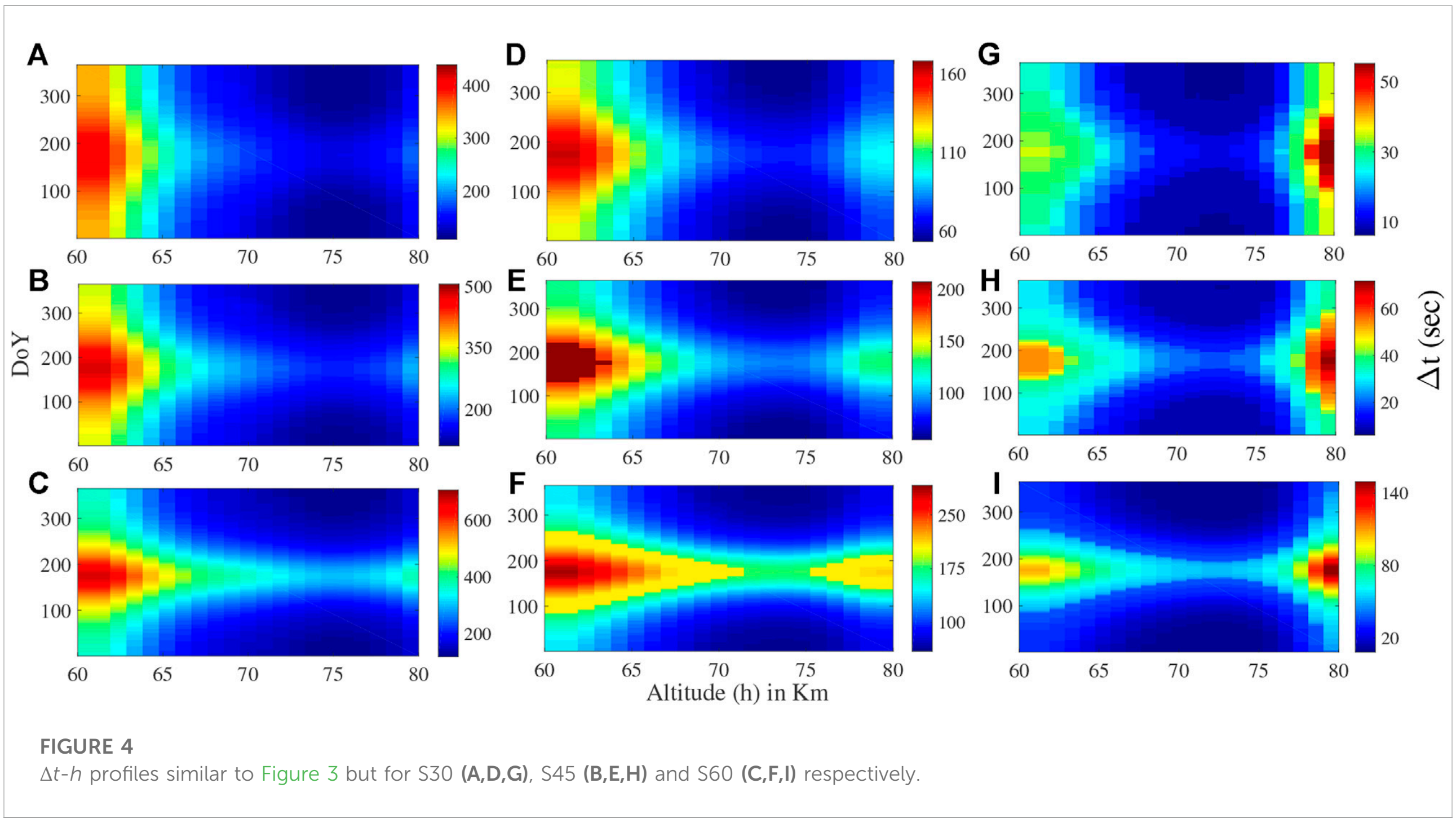

FIGURE 4
$\Delta t$-$h$ profiles similar to Figure 3 but for S30 **(A,D,G)**, S45 **(B,E,H)** and S60 **(C,F,I)** respectively.

## 3 Results and discussions

In this work, we numerically solve the D-region "electron continuity equation" (Eq. 2) separately at each of the altitudes ($h$) ranging from 60 km to 80 km and get $N_e(t)$ and subsequently $\Delta t$ at different altitudes (from Eq. 1) for the C5.2, M5.2, and X2.2 classes of flares. The '$\chi(t)$' along with the D-region parameters play the key role in this rigorous analysis. The central theme of this analysis is divided into two parts. Firstly, we analyse the latitudinal variation of the $\Delta t$-$h$ profile within the latitude range from 30° to 60° in both the hemispheres. Secondly, we analyse the seasonal variation of the $\Delta t$-$h$ profile by assuming that the particular solar flares repeat themselves on each day of the year (*DoY*). In this case, we select some specific latitudes (Section 2) for analysing the same.

We report a significant amount of $\Delta t$ variation with D-region altitude ($h$) for the entire range of latitudes and for all *DoY*'s of our interest. In Figures 1A–C, we plot the temporal profiles of soft x-ray ($\phi(t)$) (red line) along with the obtained $N_e(t)$'s over N60 at different $h$, namely, 60 km, 65 km, 70 km, 75 km and 80 km each for C5.2, M5.2, and X2.2 classes respectively. We join $t_{N_{e,\max}}$'s at different $h$ with lines and note that $\Delta t$ varies systematically with $h$ for C5.2, M5.2, and X2.2 classes each. For Figure 1, we especially choose N60 because, for a solar flare took place around February, the typical $\Delta t$ values over N60 are comparatively greater than the same over other chosen latitudes, such as, N45, N30, S30 etc., Chakraborty and Basak (2020). Obviously, we can have $N_e(t)$'s similar to Figure 1 for all other latitudes in principle and obtain $\Delta t$'s. Palit et al. (2015) computed parameters similar to Figure 1, but for a totally different set of solar flares. They performed it by analysing sub-ionospheric Very Low Frequency (VLF) signal propagation effects over D-region numerically. Their analysis was limited to the D-region within the vicinity of the VLF signal propagation path, i.e., nearly across 22°27′ N, 87°4′ E. The order of the values of $N_e(t)$ and $\Delta t$ obtained by Palit et al. (2015) and the same obtained from our investigation are similar, which does a first order validation of our analysis. For example, $\Delta t \sim 180$ s for a C-class flare at $h = 65$ km as reported by Palit et al. (2015) and same for us is 221.2 s over N30. Again, $\Delta t \sim 10$ s for an X-class flare at $h = 70$ km for them and the same for us is 10.2 s over N30. Furthermore, their model explored the D-region in the neighbourhood of the VLF signal propagation path, whereas, our method is capable to inspect anywhere within the mid-latitude D-region.

Now, we report the outcomes from Figure 1. 1) For any particular solar flare, $N_e(t)$ increases almost exponentially with $h$. Thomson, 1993 reported similar exponential increment of $N_e(t)$ with altitude. For example, $N_{e,max}$'s at $h = 65$, 70 and 75 km are $1.3 \times 10^9$ m$^{-3}$, $2 \times 10^9$ m$^{-3}$ and $2.7 \times 10^9$ m$^{-3}$ respectively for C5.2 class. The same for X2.2 class are $9.2 \times 10^9$ m$^{-3}$, $1.3 \times 10^{10}$ m$^{-3}$ and $2.4 \times 10^{10}$ m$^{-3}$ respectively. Very strong solar flares, like, X2.2 or M5.2 classes induce a greater degree of ionisation due to the chemical dynamics involving more energy at higher D-region altitudes than the C5.2 class does. Besides, the $\alpha_{eff}$ values for an X-class flare also get lowered at higher altitudes than the other two classes of flares Palit et al. (2015). Such behavior of $N_e(t)$ is responsible for different values of $t_{N_{e,max}}$ (and $\Delta t$ subsequently) at different $h$. For example, $t_{N_{e,max}}$'s at $h$ = 65 and 75 km are 21.67 h and 21.65 h respectively for C5.2 class. 2) The range of variation of $\Delta t$ values is maximum for C5.2 class and minimum for X2.2 class. For example, $\Delta t$'s are 514.1 s, 321.6 s and 221.2 s at $h = 60$, 65 and 70 km respectively for C5.2 class. The same for the X2.2 class are 61.4 s, 36.8 s and 28.7 s respectively. Additionally, $\Delta t$ for X2.2 remains close to 28.7 s for $h$ varying within 69 and 74 km. Such tendency of $\Delta t$-$h$ profile from C to X classes are in line with the results given by Mitra and Rowe (1974), Žigman et al. (2007), Basak and Chakrabarti (2013), Palit et al. (2015), Chakraborty and Basak (2020) etc., 3) Initially, the $\Delta t$ gets decremented with increasing $h$ starting from 60 km for all three classes of flares. But after a certain $h$, the tendency becomes opposite and $\Delta t$ increases. The particular $h$ around which the $\Delta t$ alters its trend is not constant for all three flares, but varies approximately within 70 and 75 km. With this basic results regarding $\Delta t$-$h$ profile, we look further into its latitudinal (Section 3.1) and seasonal (Section 3.2) dependencies. The average $\Delta t$ value for nearly C5 class solar flares, as measured by Chakraborty et al. (2022) is 150–250 s for N60 latitude. These also are in the range of our analysis for the similar altitude, latitude and DoY.

## 3.1 Latitudinal variation of $\Delta t$-$h$ profile

We report a notable and orderly latitudinal variation of $\Delta t$-$h$ profile starting from 30° to 60° over both the hemispheres (Figure 2) as computed from Eq. 2 using appropriate set of $\chi(t)$.

Firstly, it is true for the entire range of mid-latitudes and for all three flares that, $\Delta t$ starts decreasing with increasing $h$ and reaches to minima at around 70–75 km. Thereafter, it increases again till $h = 80$ km (Figure 2). As we move both from N30 to N60 and from S30 to S60, the $\Delta t$ values increase. For example, in case of the C5.2 class, $\Delta t$ values at 70 km over N30, N45 and N60 are 153.6 s, 172.1 s and 221.2 s respectively. The slant nature of solar radiation due to comparatively higher values of $\chi(t)$ towards higher latitudes is a possible reason for such higher $\Delta t$ values Chakraborty and Basak (2020). However, such variation of $\Delta t$ with latitude-altitude is much less for X2.2 than M5.2 or C5.2 classes (Figures 2E,F). Mitra and Rowe (1972) reported that, $q(t_{N_{e,max}})/N_{e,max}^2 \sim 3 \times 10^{-6}$ cm$^3$ sec$^{-1}$ for a strong solar flare and hardly varies within 70 km and 80 km. The $q(t_{N_{e,max}})/N_{e,max}^2$ physically represents $\alpha_{eff}(t = t_{N_{e,max}})$, which plays a crucial contribution on $\Delta t$. It is a possible explanation for the limited range $\Delta t$ variation for X2.2 class in our case. Secondly, the nature of latitude dependencies of $\Delta t$-$h$ profile are not identical in both the hemispheres. On the contrary, it is in general lesser over the similar latitude in the southern hemisphere than Northern one (for the flares occurred in February) Chakraborty and Basak (2020). For example, $\Delta t$'s for the C5.2 class at N45 and S45 are 444.4 s and 366.6 s at $h = 60$ km. Thirdly, in case of C5.2 and M5.2 classes, maximum value of $\Delta t$ belongs to higher latitude and lower $h$ (Figures 2A–D). Such $\Delta t$ values for C5.2 class flare over N60 and S60 at $h = 60$ km are 514.1 s and 372.8 s respectively. The same for M5.2 class are 217.1 s and 155.7 s respectively. But for X2.2 class, the same is maximum at higher latitude and higher $h$. For example, the $\Delta t$ values at 80 km are 81.9 s and 34.8 s respectively over N60 and S60. No $\Delta t$ within our selected latitude-altitude window is greater than this for X2.2.

## 3.2 Seasonal variation of $\Delta t$-$h$ profile

We start with an interesting consideration to check the seasonal variation that each of the three solar flares repeat themselves on each day of a year ($DoY$) and we report that $\Delta t$-$h$ profile varies consistently with $DoY$ (Figures 3, 4). Chakraborty and Basak (2020) reported similar $\Delta t$-$DoY$ profile, but without any altitude information of $\Delta t$. Firstly, we report that, for any fixed altitude ($h$) in northern hemispheric latitudes (Figure 3), the $\Delta t$ starts decaying since the beginning of the year (i.e., $DoY = 1$ onward) and reaches to a minimum value during the middle of the year (i.e., $DoY = 183$ onward). Thereafter, it increases till the end of the year (i.e., $DoY = 365$). For example, over N45 latitude (Figure 3E) at $h$ = 65 km, $\Delta t$'s are 151.6 s, 90.2 s and 151.6 s for $DoY$ = 1, 183 and 365 respectively for the M5.2 class. The behaviour of $\Delta t$ is exactly opposite in southern hemispheric latitudes (Figure 4). There, the maximum $\Delta t$ occurs during the middle of the year for any $h$. For example, over S45 latitude (Figure 4H) at $h$ = 65 km, $\Delta t$'s are 14.3 s, 32.7 s and 14.3 s for $DoY$ = 1, 183 and 365 respectively for the M5.2 class. Secondly, $\Delta t$ (= 149.5 s) at $h$ = 70 km for C5.2 remains unchanged only for 30 days (i.e., $DoY$ = 157–187) over N60 (Figure 3A). But over N30 (Figure 3C), $\Delta t$ (= 137.2 s) remains unchanged for 88 days (i.e., $DoY$ = 127–215). So, after inspecting several other examples, we report that $\Delta t$ almost remains unchanged for a longer period of time comparatively over lower latitudes. Thirdly, $\Delta t$ around $h$ = 60 km is greater than the same around $h$ = 80 km both for C5.2 and M5.2 classes almost for all $DoY$'s. But, the opposite happens for X2.2 class. For example, it is 501.8 s and 223.3 s respectively at $h$ = 60 km and $h$ = 80 km on $DoY$ = 1 over N45 for C5.2 class (Figure 3B). On the other hand, it is 55.3 s and 69.6 s respectively at $h$ = 60 km and $h$ = 80 for X2.2 class (Figure 3H). Overall, there is a contrasting behaviour in case of the X2.2 class solar flare, compared to the C5.2 and M5.2 class of solar flares.

Additionally, we analyze two more solar flares, namely, C1.8 and C9.6 classes (both of them occurred on 21 January 2017) to reaffirm the validity of the analysis (Supplementary Material S1). These flares are much different in strength than the same used in the manuscript and that is the reason for choosing them here. We report that the study of $\Delta t$-$h$ profile and its latitude dependence for this two solar flares are in agreement of the similar analysis done. Temporal profiles of solar x-ray flux ($\phi(t)$) during C1.8 and C9.6 classes of solar flares and corresponding D-region electron density profiles ($N_e(t)$) at different altitudes (60 km, 65 km, 70 km, 75 km, and 80 km respectively) over 60° north latitude (N60) are computed (Figure 1; Supplementary Material S1). $\Delta t$ values at different altitudes for C1.8 class of flare are 530 s, 528 s, 450 s, 405 s, and 528 s respectively. We show the latitudinal variation of $\Delta t$-$h$ profile for this two solar flares over all the latitudes starting from 30° to 60° (Figure 2; Supplementary Material S1). In case of seasonal variation of $\Delta t$-$h$ profile for the solar flares, we report similarity in overall trend between the results presented in Figures 3, 4 (Supplementary Material S1) and here. Now, we provide the following explanations. Palit et al. (2013) reported that, the maximum electron production rate ($q_{max}$) per unit volume during peak of a M-class solar flare is noted at nearly $h$ = 90 km. This altitude value is well beyond the range of our computation. Whereas, for X-class solar flare $q_{\max}$ is noted at $h$ = 81 km, which is almost close to our upper limit of $h$. Mitra and Rowe (1972) reported that for a strong solar flare (likely a X-class solar flare) the rate of decay of $\alpha_{eff}$ (at $t_{N_{e,max}}$) and $\lambda$ with $q(t)$ is exactly the same. Whereas, the rates are not same for the other two classes of solar flares. Since, $q(t)$ has a strong altitude dependency ($q(t)$ increases with altitude), the natures of $\alpha_{eff}$ (at $t_{N_{e,max}}$) and $\lambda$ may be responsible for different cases observed in X-class.

# 4 Conclusion

In this work, we dealt with the altitude ($h$) dependency of D-region response time delay ($\Delta t$) to an incoming solar irradiation during solar flare. We did it for three solar flares of different classes, namely, C5.2, M5.2, and X2.2 (Table 1). Alongside, we did similar treatment with two other solar flares, namely, C1.8 and C9.6 classes (Supplementary Material S1). They have occurred during the month of February of different years. We obtained their $\Delta t$-$h$ profiles (Figures 1–4) by solving the D-region "electron continuity equation" (Eq. 2), where, $\chi$ estimates the variation of solar irradiation onto the ionosphere. We established $\Delta t$-$h$ profile and its variation with latitude (Figure 2) and $DoY$ (Figures 3, 4) successfully. Firstly, from the $N_e(t)$ profiles are different D-region altitudes ($h$), we confirmed the $h$ dependency of $\Delta t$ (Figure 1) for all of the flares. We noted that the $\Delta t$ decreases with increasing $h$ initially, and then it increases (Figure 1) after a certain value of $h$. Necessary reason for it is the altitude profile of $q(t)$ and the chemical properties of the D-region Mitra and Rowe (1972). Secondly, we got a significant latitudinal variation of the $\Delta t$-$h$ profile. We noted that, as a whole the $\Delta t$'s (for a fixed $h$) have lesser values in Southern hemispheric latitude than that of northern hemisphere for any given solar flare (Figure 2). The possible reason is the time of occurrences of these solar flares are in the month of February, i.e., during winter in northern hemisphere (and summer in the other). Hence, the solar irradiation over the northern hemisphere is more slant, i.e., the $\chi$ values are higher. As a result, the effective solar irradiation intensity gets lowered comparatively. Higher values of $\chi$ leads to greater values of $\Delta t$ Chakraborty and Basak (2020) and *vice versa*. Likewise, in case of seasonal variation, $\chi$'s are relatively less during middle of the year ($DoY \sim 170$–190) compared to the other times of the year in northern hemisphere (Figure 3). In Southern hemisphere, $\chi$ profile is exactly opposite (Figure 4). It explains the possible reason behind the relatively lower values of $\Delta t$ in middle of the year over Northern hemisphere. Thirdly, we observed an contrasting behaviour in case of X2.2 class compared to C5.2 and M5.2 classes pertain to both latitude and seasonal effects on $\Delta t$-$h$ profile (see Section 3.1, Section 3.2). We reaffirm that the entire methodology stated here is not limited to investigate the $\Delta t$-$h$ profiles for three solar flares analysed here, but the C1.8 and C9.6 classes of flares show similar results (Supplementary Material S1). The overall applicability of it covers from C1.8 to X2.2 class. Hence, it is applicable to the entire range of available solar flares.

## Data availability statement

The datasets presented in this study can be found in online repositories. The solar X-ray and the solar zenith data is available from https://www.ngdc.noaa.gov/ and https://www.esrl.noaa.gov/ respectively.

## Author contributions

SC and TB contributed to conception and design of the analysis. TB organised the satellite data and solar zenith angle profile. SC performed the computation and wrote the first draft of the manuscript. TB, SC, and RP contributed to the manuscript revision with scientific input. All the authors approved the submitted manuscript.

## Acknowledgments

Authors thank NCEI-NOAA for using solar X-ray data and solar zenith angle profile. SC acknowledges the support of DST-INSPIRE fellowship, Department of Science and Technology, India (Application Reference No. DST/INSPIRE/03/2021/001103; IF No. IF200266).

## Conflict of interest

The authors declare that the research was conducted in the absence of any commercial or financial relationships that could be construed as a potential conflict of interest.

## Publisher's note

All claims expressed in this article are solely those of the authors and do not necessarily represent those of their affiliated organizations, or those of the publisher, the editors and the reviewers. Any product that may be evaluated in this article, or claim that may be made by its manufacturer, is not guaranteed or endorsed by the publisher.

## Supplementary material

The Supplementary Material for this article can be found online at: https://www.frontiersin.org/articles/10.3389/fenvs.2022.1020137/full#supplementary-material